**Journal**
## Perspective

# How the Brain Might Work:
# Statistics Flowing in Redundant Population Codes

Xaq Pitkow[1,2] and Dora E Angelaki[1,2]
[1] Department of Neuroscience, Baylor College of Medicine
[2] Department of Electrical and Computer Engineering, Rice University

**It is widely believed that the brain performs approximate probabilistic inference to estimate causal variables in the world from ambiguous sensory data. To understand these computations, we need to analyze how information is represented and transformed by the actions of nonlinear recurrent neural networks. We propose that these probabilistic computations function by a message-passing algorithm operating at the level of redundant neural populations. To explain this framework, we review its underlying concepts, including graphical models, sufficient statistics, and message-passing, and then describe how these concepts could be implemented by recurrently connected probabilistic population codes. The relevant information flow in these networks will be most interpretable at the population level, particularly for redundant neural codes. We therefore outline a general approach to identify the essential features of a neural message-passing algorithm. Finally, we argue that to reveal the most important aspects of these neural computations, we must study large-scale activity patterns during moderately complex, naturalistic behaviors.**

## INTRODUCTION
### Perception as inference

In its purest form, probabilistic inference is the 'right' way to solve problems (Laplace 1812). While animal brains face various constraints and cannot always solve problems in the smartest possible way, many human and animal behaviors do provide strong evidence of probabilistic computation. The idea that the brain performs statistical inference harkens back at least to Helmholtz (1925) and has been elaborated by many others. According to this hypothesis, the goal of sensory processing is to identify properties of the world based on ambiguous sensory evidence. Since the true properties cannot be determined with perfect confidence, there is a probability distribution associated with different interpretations, and animals weigh these probabilities when choosing actions. It is widely accepted that the brain somehow approximates probabilistic inference. But how does the brain do it?

To understand the neural basis of the brain's probabilistic computations, we need to understand the overlapping processes of encoding, recoding, and decoding. *Encoding* describes the relationship between sensory stimuli and neural activity patterns. *Recoding* describes the dynamic transformation of those patterns into other patterns. *Decoding* describes the use of neural activity to generate actions.

In this paper we speculate how these processes might work, and what can be done to test it. Our model is consistent with the widely embraced principle that the brain performs probabilistic inference, but we offer a hypothesis about how those computations could appear within dynamic, recurrently connected probabilistic population codes to implement a class of algorithms called message-passing algorithms. This way of thinking about neural representations allows one to connect the computations operating on interpretable world variables to the computations operating on neural activations. The connection between these levels is made by summary statistics, which are functions of the neural data that encode the parameters of the inference algorithm. It is these summary statistics that we describe in the title as flowing through the brain. To explain this model, we review and connect some foundational ideas from statistics, probabilistic models, message-passing, and population codes.

We argue that to reveal the structure of these computations, we must study large-scale activity patterns in the brains of animals performing naturalistic tasks of greater complexity than most current neuroscience efforts. We also emphasize that since there are many equivalent ways for the brain to implement natural computations (Marder and Taylor 2011), one can understand them best at the representational level (Marr 1982) — characterizing how statistics encoding *task variables* are transformed by neural computations — rather than by fine details of how the large-scale neural activity patterns are transformed.

### *A critique of simple tasks*

Neuroscience has learned an enormous amount in the past several decades using a simple kind of tasks, such as those in which subjects choose between two options — two-alternative forced-choice tasks (2AFC). Many experiments found individual neurons that were tuned to task stimuli with enough reliability that only a small handful could be averaged together to perform as well or better than the animal in a 2AFC task (Newsome et al. 1989; Cohen & Newsome 2009; Gu et al. 2008; Chen et al. 2013a). With a brain full of neurons, why isn't behavior better? One possible answer is that responses to a fixed stimulus are correlated. These covariations cannot be averaged away by pooling, limiting the information the animal has about the world and creating a redundant neural code (Zohary et al. 1994, Moreno-Bote et al. 2014). Individual neural responses also correlate with reported percepts, even when the stimulus itself is perfectly ambiguous. This has been reported in multiple tasks and cortical areas (Britten et al. 1996; Uka & DeAngelis 2004; Nienborg & Cumming 2007; Gu et al. 2008; Fetsch et al. 2011;

Chen et al. 2013a,b,c; Liu et al. 2013a,b). How much of these correlations arise from feedforward sensory noise versus feedback remains unresolved, in part because the tasks in which they are measured do not readily distinguish the two. This is one example of how simple tasks can create problems.

The most fundamental problem is that simple tasks limit the computations and neural activity to a domain where the true power and adaptability of the brain is hidden. When the tasks are low-dimensional, the mean neural population dynamics are bound to a low-dimensional subspace, and measured mean neural activity seems to hit this bound (Gao and Ganguli 2015). This means that the low-dimensional signals observed in the brain may be an artifact of overly simple tasks. Even worse, many of our standard tasks are linearly solvable using trivial transformations of sense data. And if natural tasks could be solved with linear computation, then we wouldn't even need a brain. We could just wire our sensors to our muscles and accomplish the same goal, because multiple linear processing steps is equivalent to a single linear processing step. Distinguishing these steps becomes difficult at best, and uninterpretable at worst.

Finally, principles that govern neural computation in overtrained animals performing unnatural lab tasks may not generalize. Are we learning about the real brain in action, or a laboratory artifact? Evolution did not optimize brains for 2AFC tasks, and the real benefit of complex inferences like weighing uncertainty may not be apparent unless the uncertainty has complex structure. How can we understand how the brain works without challenging it with tasks like those for which it evolved? We return to the question of task design later, after we have introduced our computational model for probabilistic inference that we want to test.

## ALGORITHM OF THE BRAIN

In perception, the quantities of interest — the things we can act upon — cannot be directly observed through our senses. These unobservable quantities are called latent or hidden variables. For example, when we reach for a mug, we never directly sense the object's three-dimensional shape — that is latent — but only images of reflected light and increased tactile pressure in some hand positions. Some latent quantities are relevant to behavioral goals, like the handle's orientation, while other latent variables are a nuisance, like shadows of other objects. Perception is hard because both types of latent variables affect sensory observations, and we must disentangle nuisance variables from our sense data to isolate the task-relevant ones (DiCarlo and Cox 2007).

We must infer all of this based on uncertain sensory evidence. There are multiple sources of uncertainty. Some is intrinsic to physics: lossy observations due to occlusion or photon shot noise. Some is unresolvable variation, like the hum of a city street. Other uncertainty is due to biology, including neural noise and limited sampling by the sensors and subsequent computation. Uncertainty also arises from suboptimal processing (Beck et al. 2012): model mismatch behaves much like structured noise. Regardless of its origin, since uncertainty is an inevitable property of perceptual systems, it is valuable to process signals in accordance with probabilistic reasoning.

Unfortunately, exact probabilistic inference is intractable for models that are as complex as the ones our brains seem to make. First, merely representing arbitrary joint probabilities exactly requires enormous resources, exponential in the number of variables. Second, performing inference over these distributions requires an exponentially large number of operations. This means that exact inference in arbitrary models is out of the question, for the brain or any other type of computer. Finally, even exploiting the structure in the natural world, a lifetime of experience never really has enough data to constrain a complete statistical model, nor do we have enough computational power and time to perform statistical inference based on these ideal statistics. Our brain must invoke the 'blessing of abstraction' (Goodman et al. 2009) to overcome this 'curse of dimensionality' (Bellman 1957). The brain must make assumptions about the world that limit what it can usefully represent, manipulate and learn — this is the 'no free lunch theorem' (Wolpert 1996).

In the next sections, we describe our neural model for encoding structured probabilities, and recoding or transforming them during inferences used to select actions.

### *Encoding: Redundant distributed representations of probabilistic graphical models*

Since the probability distribution over things in the world is hugely complex, we hypothesize that the brain simplifies the world by assuming that not every variable necessarily interacts with all other variables. Instead, there may be a small number of important interactions. Variables and their interactions can be elegantly visualized as a sparsely connected graph (Figure 1A), and described mathematically as a probabilistic graphical model (Koller and Friedman 2009). These are representations of complex probability distributions as products of lower-dimensional functions (see Box). Such constraints on possible distributions are appropriate for the natural world, which has both hierarchical and local structures.

Knowledge about the world is embodied in these interactions between variables. Many of the most important ones express nonlinear relationships between variables. For example, unlike pervasive models of sparse image coding (Olshausen and Field 1997), natural images are not generated as a simple linear sum of features. Instead many properties of natural images arise from occlusion (Pitkow 2010) and the multiplicative absorption of light (Wainwright and Simoncelli 2000).

How might these probabilistic graphical models be represented by neural activity? Information about each sensory variable is spread across spiking activity of many neurons with similar stimulus sensitivities. Conversely, neurons are also tuned to many different features of the world (Rigotti et al. 2013). Together, these facts mean that the brain uses a distributed and multiplexed code.

There are several competing models of how neural populations encode probabilities, with different advantages and disadvantages, and accounting for different aspects of experimental observations. In temporal representations of uncertainty, like the 'sampling hypothesis,' instantaneous neural activity represents a single interpretation, without uncertainty, and probabilities are reflected by the set of interpretations over time (Hoyer and Hyvärinen 2003; Berkes et al. 2011; Moreno-Bote et al. 2011; Buesing et al. 2011; Haefner et al. 2016; Orbán et al. 2016).

Here we focus on spatial representations of probability. According to these models, the spatial pattern of neural activity across a population of neurons implicitly encodes a probability distribution (Ma et al. 2006; Jazayeri and Movshon 2006; Savin and Denève 2014; Rao 2004; Hoyer and Hyvarinen 2003). We consider, specifically, a multivariate elaboration of probabilistic population codes (PPCs, Ma et al. 2006). In a simple example, the population vector encodes the best point estimate of a stimulus variable, and the total spike count in a population can reflect the overall confidence about the variables of interest (Ma et al. 2006). More generally, every neural spike adds log-probability to some interpretations of a scene, so more spikes typically means more confidence. This class of models has been used successfully to explain the brain's representation of uncertainty (Ma et al. 2006; Jazayeri and Movshon 2006), evidence integration over time (Beck et al. 2008), multimodal sensory integration (Fetsch et al. 2012), and marginalization (Beck et al. 2011).

Past work has critiqued this approach as generally requiring an inordinate number of neurons (Fiser et al. 2010). If one wanted to represent a univariate probability with a resolution around 1/10 of the full variable range, one would need 10 neurons. But to represent an arbitrary joint probability distribution over a mere 100 variables to the same resolution along *each* dimension, this number explodes to $10^{100}$ neurons, more than the number of particles in the visible universe.

Thankfully, the brain does not have to represent arbitrary probabilities in an unstructured world. By modeling the structure of the world using probabilistic graphical models, the required resources can be diminished enormously, requiring only enough neurons to describe the variables' direct *interactions* (see Box, Figure 1A) with the desired precision. To represent a joint distribution over the same 100 variables with sparse direct pairwise interactions (10% of possible pairs), we only need a reasonable $10^4$ neurons, saving an impressive 96 orders of magnitude. How the brain learns good models is a fundamental open question, but one we do not address here. Instead we consider the computations that the brain could perform using an internal model that has already been constructed.

We therefore hypothesize that the brain represents posterior probabilities by a probabilistic graphical model, implemented in a multivariate probabilistic population code. In this model, estimates and uncertainties about both individual latent variables and groups of variables are represented as

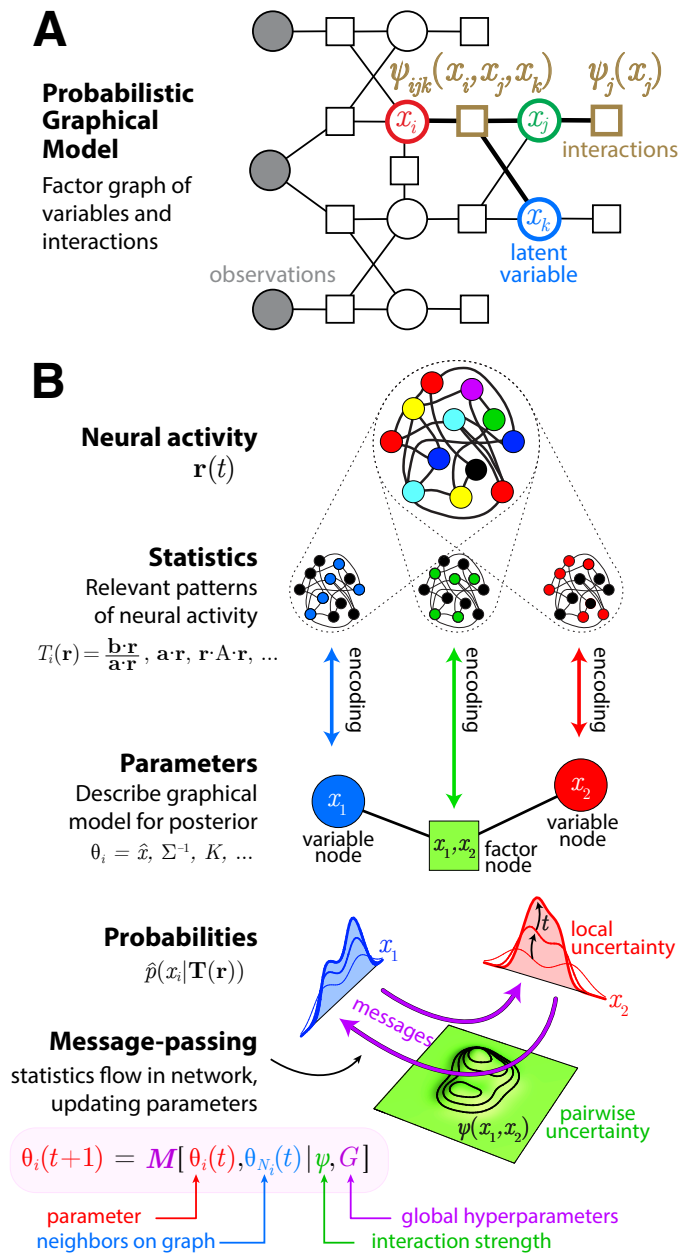

**Figure 1. Inference in graphical models embedded in neural activity.**

**A**. A probabilistic graphical model compactly represents direct relationships (statistical interactions or conditional dependencies, see Box) amongst many variables. This can be depicted as a factor graph, with circles showing variables $x$ and squares showing interactions $\psi$ between subsets of variables. Three variable nodes (red, green, blue) and two interaction factors (brown) are highlighted. Observed variables are shaded gray. **B**. Illustration of our message-passing neural inference model. Neural activity resides in a high-dimensional space $\mathbf{r}$ (top). Distinct statistics $T_i(\mathbf{r})$ reflect the strength of certain patterns within the neural activity (red, green, and blue uniformly colored activation patterns). These statistics serve as estimators of parameters $\theta_i$, such as the mode or inverse covariance, for the brain's approximate posterior probabilities over latent variables (colored probability distributions). As neuronal activities evolve over time, the specific patterns that serve as summary statistics are updated as well. We hypothesize that the resultant recoding dynamics obey a canonical message-passing equation that has the same form for all interacting variables. This generic equation updates the posterior parameters at time $t$ according to $\theta_i(t+1) = M[\theta_i(t), \theta_{N_i}(t) ; \psi, G]$, where $M$ represents a family of message-passing functions, $\theta_{N_i}$ represents all parameters in the the set of neighbors $N_i$ on the graphical model that interact with $\theta_i$, the factors $\psi$ determine the particular interaction strengths between groups of variables, and $G$ are global hyperparameters that specify one message-passing algorithm from within a given family. The message-passing parameter updates can be expressed equivalently in terms of the neural statistics since $T_i(\mathbf{r},t) \approx \theta_i(t)$. The statistics flow through the population codes as the neural activity evolves, thereby representing the dynamics of probabilistic inference on a graphical model.

---

**Probabilistic inference and Population codes**

<u>Probabilistic inference</u>: Drawing conclusions based on ambiguous observations. Typical inference problems include finding the marginal probability of a task-relevant variable, or finding the most probable explanation of observed data.

<u>Probabilistic computation</u>: Transformation of signals in a manner consistent with rules of probability and statistics, especially through appropriate sensitivity to uncertainty (Ma 2012).

<u>Latent variables</u> (also called hidden or causal variables): Quantities whose value cannot be directly observed, yet which determine observations. Latent variables may be *relevant* or *irrelevant* (*nuisance*), depending on the task. Latent variables are typically the target of inferences and actions.

<u>Probabilistic graphical model</u>: A decomposition of a probability distribution into a product of conditionally independent factors that each describe interactions between distinct subsets of variables. One useful such model is a factor graph (Figure 1A) that represents a structured probability distribution $P(\mathbf{x}_\alpha) = \prod_\alpha \psi_\alpha(\mathbf{x}_\alpha)$ where $\mathbf{x}=(x_1,\ldots,x_n)$ is a vector of all variables and $\mathbf{x}_\alpha$ is a subset of variables that interact through the function, or factor, $\psi_\alpha(\mathbf{x}_\alpha)$.

<u>Statistical interaction</u>: Dependency between two variables that cannot be explained by other observed covariates. These are often represented by squares in a factor graph. Statistical interactions may be generated by physical interactions in the world, or can arise due to some neglected latent variables.

<u>Higher-order interaction</u>: A nonlinear statistical interaction between variables. An especially interesting case is when three or more variables interact. This leads naturally to contextual gating, whereby one variable (the 'context') determines whether two others interact.

<u>Parameters</u>: Quantities $\theta_i$ that determine a probability distribution. For example, the mean and variance determine a specific Gaussian distribution. Here we consider parameters of the posterior distribution over latent variables.

<u>Statistic</u>: A function $T_i(\mathbf{r})$ of observable data (e.g., for the brain, the neural activity $\mathbf{r}$) that can be used to infer a parameter. If the statistic contains the same information about a parameter as all observations together, the statistic is called 'sufficient'. For example, the total spike count $\mathbf{1} \cdot \mathbf{r}$ in a model population could be a statistic reflecting the confidence (1/variance) about a latent variable. (Note for aficionados: frequentists use statistics to summarize the probability of observed data. We are using the Bayesian version, where the statistics describe the probability of latent variables that generated the data.)

<u>Message-passing algorithm</u>: An iterative sequence of computations that performs a global computation by operating locally on statistical information ('messages') conveyed exclusively along the edges of a probabilistic graphical model.

<u>Population code</u>: Representation of a sensory, motor, or latent variable by the collective activity of many neurons.

<u>Probabilistic population code</u>: A population code that represents a probability distribution. This requires at least two statistics, to encode at least two parameters, such as an estimate and uncertainty.

<u>Information-limiting correlations</u> (informally, 'bad noise'): Covarying noise fluctuations in large populations that are indistinguishable from changes in the encoded variable. These arise when sensory signals are embedded in a higher-dimensional space, or when suboptimal upstream processing throws away extensive amounts of information. These noise correlations cannot be averaged away by adding more neurons (Moreno-Bote et al. 2014).

<u>Redundancy</u>: Identical information content in different signals. If two neuronal populations inherit the same limited information from an upstream source, then either population can be decoded separately, or the two can be averaged, and the result is the same.

<u>Robustness</u>: Insensitivity of outputs to variations in the network. A computation is robust whenever uncertainty added by suboptimal processing is much smaller than the intrinsic uncertainty caused by information-limiting noise.

---

distinct but possibly overlapping spatial patterns of neural activity (Figure 1B, Raju and Pitkow 2016). These activity patterns serve as 'summary statistics' (see Box) that tell us about different properties or parameters of the posterior distribution.

Not every aspect of the neural responses will be a summary statistic encoding information about latent variables. The other aspects would count as irrelevant noise. In quantifying the neural code, we can then compress the neural responses, aiming to keep those aspects that are relevant to the brain's posterior probability. If this compression happens to be lossless, then these summary statistics are called 'sufficient statistics.'

This concept of statistics is similar to the familiar notions of time- or trial-averaged means and covariances, except they are potentially more general functions of neural activity and estimate a wider range of parameters about the posterior probabilities. They are more like a single-trial population average — in fact for a population of homogeneous poisson neurons with gaussian tuning curves, the population sum is proportional to the inverse variance of the posterior (more spikes, more certainty). For *linear* probabilistic population codes, the sufficient statistics are weighted averages of neural activity (Ma et al. 2006). More generally the statistics may be arbitrary nonlinear functions, depending on how the relevant latent variables influence the neurons being considered.

Collectively, these summary statistics are a dimensionality reduction that encodes the brain's probabilistic knowledge. This reduction enables us to relate the transformations of neural activity to the transformations of the posterior probabilities during inference.

### *Recoding: nonlinear transformations*

The brain does not simply encode information and leave it untouched. Instead, it recodes or transforms this information to make it more useful. It is crucial that this recoding is nonlinear, for two reasons: first, for isolating task-relevant latent variables, and second, for updating parameters for probabilistic inference.

In the first case, nonlinear computation is needed to extract task-relevant variables, because in natural tasks they are usually nonlinearly entangled with task-irrelevant 'nuisance' variables (DiCarlo and Cox 2007). This is true even when there is no uncertainty about the underlying variables. Figure 2A illustrates how nonlinear computations can allow subsequent linear computation to isolate task variables.

Ethological tasks typically require complex nonlinearities to untangle task-relevant properties from nuisance variables. In principle, any untangling can be accomplished by a simple network with one layer of nonlinearities, since this architecture is a universal function approximator for both feedforward nets (Cybenko 1989; Hornik 1991) and recurrent nets (Schäfer and Zimmerman 2007). However, in practice this can be more easily accomplished by a 'deep' cascade of simpler nonlinear transformations. This may be because the parameters are easier to learn, because the hierarchical structure imposed by the deep model are better matched to natural inputs (Montúfar et al. 2014), because certain representations use brain resources more economically, or all of the above. Indeed, trained artificial deep neural networks have notable similarities with biological neural networks (Yamins et al. 2014).

To what extent are these nonlinear networks probabilistic? One can trivially interpret neural responses as encoding probabilities, because neuronal responses differ upon repeated presentations of the same stimulus, and according to Bayes' rule that means that any given neural response could arise from multiple different stimuli. But to actually *use* that encoding of probability to perform inference (exactly or approximately), the brain's decisions must be influenced by uncertainty, such as placing less weight on less reliable information. Crucially, reliability often varies over time, so good inferences and the brain computations that produce them should be sensitive to the current uncertainty, rather than merely the typical uncertainty (Ma 2012).

Using probabilistic relationships typically requires a second kind of nonlinear transformation, one that accounts not only for relations between latent variables, but also between parameters of the posteriors about them, or equivalently between the sufficient statistics of the neural responses. Some of these parameter updates can be linear: to accumulate information about a latent variable the brain can just count spikes over time (Beck et al. 2008). Others will be nonlinear even in simple cases. For example, in a gaussian graphical model, the latent variables themselves are linearly correlated; but even in this simple case, the parameters describing uncertainties about the variables require nonlinear updates during inference. This is because the inferred mean of one gaussian variable based on all available evidence is a complex function of the means and variances of other variables.

Neural recoding should thus account for nonlinear relationships both between latent variables and between posterior parameters reflecting uncertainties. These nonlinear computations should collectively approximate probabilistic inference.

### Recurrent recoding: Inference by message-passing

Exact inference in probabilistic models is intractable except in rare special cases. Many algorithms for approximate inference in probabilistic graphical models are based on iteratively transmitting information about probability distributions along the graph of interactions. These algorithms go by the name of 'message-passing' algorithms, because the information they convey between nodes are described as messages. They are dynamical systems whose variables represent properties of a probability distribution. Message-passing is a broad class of algorithms that includes belief propagation (Pearl 1988), expectation propagation (Minka 2001), mean-field inference, and other types of variational inference (Wainwright and Jordan 2008). Even some forms of sampling (Geman and Geman 1984; Lee and Mumford 2003) can be viewed as message-passing algorithms with a stochastic component.

Each specific algorithm is defined by how incoming information is combined and how outgoing information is selected. To be a well-defined message-passing algorithm, these operations must be the same, irrespective of what kinds of latent variables are being inferred, or how strongly they interact in the underlying probabilistic graphical model.

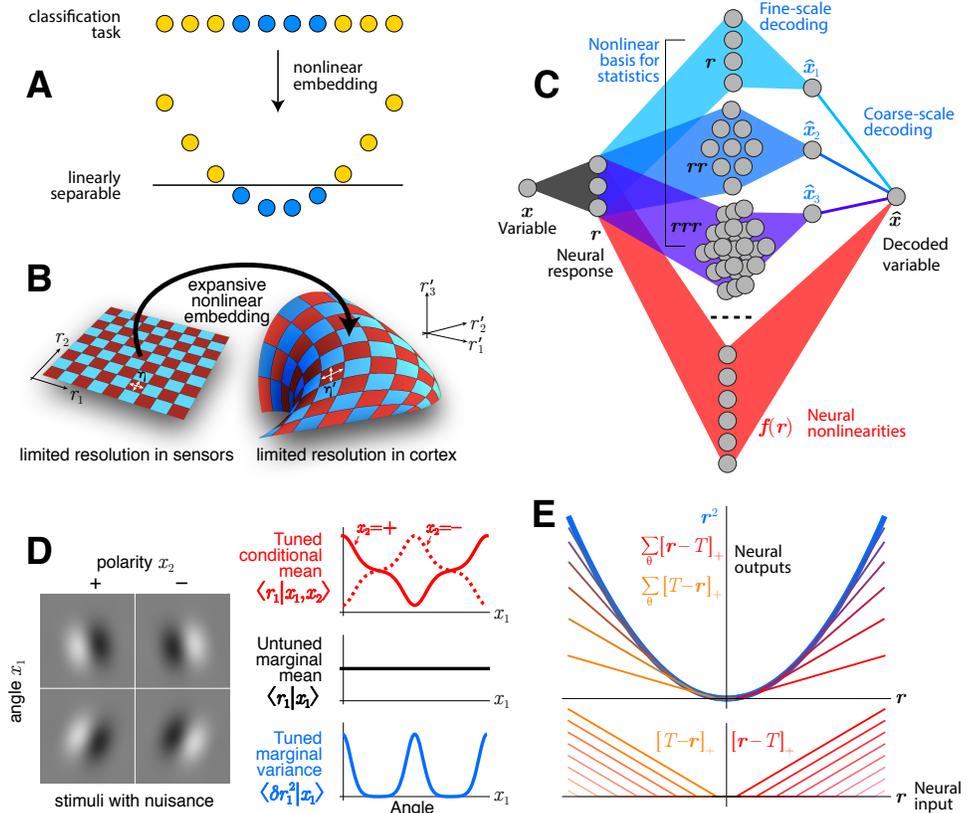

**Figure 2. Nonlinearity and redundancy in neural computation**
**A**: The task of separating yellow from blue dots cannot be accomplished by linear operations, because the task-relevant variable (color) is entangled with a nuisance variable (horizontal position). After embedding the data nonlinearly into a higher-dimensional space, the task-relevant variable becomes linearly separable. **B**: Signals from the sensory periphery have limited information content, illustrated here by a cartoon tiling of an abstract neural response space $r = (r_1, r_2)$. Each square represents the resolution at which the responses can be reliably discriminated, up to a precision determined by noise ($\eta'$). When sensory signals are embedded into a higher-dimensional space $r' = (r'_1, r'_2, r'_3)$, the noise is transformed the same way as the signals. This produces a redundant population code with high-order correlations. **C**: Such a code is redundant, with many copies of the same information in different neural response dimensions. Consequently, there are many ways to accomplish a given nonlinear transformation of the encoded variable, and thus not all details about neural transformations (red) matter. For this reason it is advantageous to model the more abstract, representational level on which the nonlinearity affects the information content. Here, a basis of simple nonlinearities (e.g. polynomials, bluish) can be a convenient representation for these latent variables, and the relative weighting of different coarse types of nonlinearities may be more revealing than fine details as long as the different readouts of fine-scale are dominated by information-limiting noise. **D**: A simple example showing that the brain needs nonlinear computation due to nuisance variation. Linear computation can discriminate angles $x_1$ of Gabor images $I$ of a fixed polarity $x_2=\pm1$, because the nuisance-conditional mean of the linear function $r=w \cdot I$ is tuned to angle (red). If the polarity is unknown, however, the mean is not tuned (black). Tuning is recovered in the variance (blue), since the two polarities cause large amplitudes (plus and minus) for some angles. Quadratic statistics $T(r)=r^2$ therefore encode the latent variable $x_1$. **E**: A quadratic transformation $r^2$ can be produced by the sum of several rectified linear functions $[\cdot]_+$ with uniformly spaced thresholds $T$.

Differences between algorithms reflect different choices of local approximations for intractable global computations. For instance, in belief propagation, each operation treats all incoming evidence as independent, even if it is not. Some randomness may be useful to represent and compute with distributions (Hinton and Sejnowski 1983; Hoyer and Hyvärinen 2003) as well as overcome blind spots in suboptimal message-passing algorithms (Pitkow et al. 2011).

The result of all of these circulating messages is that all evidence indirectly related to each variable is eventually consolidated in one place on the probabilistic graphical model, so any inferred variable can be read out directly (Raju and Pitkow 2016).

This is a broad family of algorithms, but not all-encompassing. In particular, message-passing restricts direct information flow only along the graphical model: if two variables are conditionally independent according to the model, then they are not directly connected on the graph, and messages cannot flow directly between them. We and many others have hypothesized that the brain uses some version of belief propagation (e.g. Lee and Mumford 2003; Raju and Pitkow 2016), which specifies a functional form by which posterior parameters interact on a probabilistic graphical model. This message-passing algorithm performs exact inference on tree graphs, and often performs well even in other graphs. Nonetheless, belief propagation has inferential blind spots (Heinemann and Globerson 2011; Pitkow et al. 2011) that the brain must overcome. We speculate that the brain has its own clever version of message-passing that can compensate for such handicaps. It will be important to define families of message-passing algorithms from which to select good fits to neural recoding. In later sections we will describe an analysis framework that may help determine which probabilistic graphical models and message-passing algorithms best match the brain's computations.

### *Decoding: Probabilistic control*

The brain appears to use multiple strategies to map its beliefs onto actions, depending partly on the task structure and the animal's ability to model it (Dolan and Dayan 2013). Two types of strategies are known as model-free control, in which an agent essentially builds a look-up table of likely rewards, and model-based control, in which an agent relies on an internal model of the world to predict future rewards.

In the latter case, choosing a good action can be formulated as an optimal control problem, where the animal aims to maximize expected utility based on its uncertain beliefs about the current and future latent variables (Sutton and Barto 1998). Maximizing utility thus involves building not only a model of the external world, but also a model of one's own causal influence on it, as well as an explicit representation of uncertainty. Probabilistic inference should therefore play a major role in guiding action. This should make our theory of statistics flowing through redundant population codes especially useful for understanding neural computation all the way from encoding, through recoding, to decoding.

## FROM THEORY TO EXPERIMENT

If our theory of neural computation is correct, how could we test it? We propose an analysis framework for quantifying how neural computation transforms information at the representational level. This framework compares encoding, recoding and decoding between a behavioral model and the neural statistics that represent the corresponding quantities. We then argue that such a description is especially natural for redundant population codes. We conclude by advocating for more complex tasks that can expose the brain's flexible computations, by including more uncertainty, latent variables, and nuisance variables.

### *Analyzing neural responses at the representational level*

The fundamental reason that we can fruitfully model computation at the representational level is that information about any given relevant latent variables is distributed amongst many neurons in the brain. We must therefore think about the basic properties of these neural populations.

In conventional population codes, the relevant signal is shared by many neurons. By combining neural responses, the brain can then average away noise. In *redundant* population codes, in contrast, the noise is shared as well. Specifically, the noise fluctuates in some patterns that are indistinguishable from the signal pattern across the population. These are 'information-limiting correlations' (Zohary et al. 1994; Moreno-Bote et al. 2014; Pitkow et al. 2015). The brain cannot average away this noise.

The disadvantage of a redundant population code is that such a neural representation encodes a limited amount of information, whether about estimates of latent variables or a posterior distribution over them. This information limit arises originally at the sensory input itself (Kanitscheider et al. 2015): once signals enter the brain, the neural representation expands massively, engaging many times more neurons than sensory receptors. Yet despite the large increase in the number of neurons, the brain cannot encode more information than it receives. So as long as all of these extra neurons are not just remembering the past (Ganguli et al. 2008), then they can at best recode the relevant signals in a new form, and may lose information due to biological constraints or suboptimal computation (Beck et al. 2012; Babadi and Sompolinsky 2014).

The advantage of a redundant population code is that multiple copies of essentially the same information exist in the network activity patterns. Decoding these patterns in many different ways therefore can produce similar outcomes (Pitkow et al. 2015), providing a measure of robustness to neural computation.

Since neural networks can implement computationally useful transformations in multiple ways, we should therefore focus on the shared properties of equivalent computations. It may not even be possible to identify the fine-scale structure of neural interactions based on reasonable amounts of data (Haefner et al. 2013). A focus on the coarse-scale properties should abstract away the fine implementation details while preserving essential properties of the nonlinear computation (Yang and Pitkow 2015; Lakshminarasimhan et al. 2017). (Marder and Taylor 2011) similarly advocate for considering equivalent classes of low-level biological mechanisms for a given circuit function, since neural circuits are degenerate and robust even down to the level of ion channel distributions.

The essential encoding properties comprise those statistics of the stimulus that are informative about a task, and the essential recoding transformations are determined by the nonlinear functions needed to untangle the nuisance variables. For example, an energy-based model for phase-invariant orientation selectivity would have quadratic computations as an essential property (Figure 2D). This quadratic computation can be performed in two equivalent ways: squaring neural responses directly, or summing a set of rectifiers with uniformly spaced thresholds (Figure 2E).

These quadratic and rectified computations are equivalent not only for transforming the signal, but also for transforming the noise that enters with the signal, particularly the information-limiting fluctuations that look indistinguishable from changes in the signal. In redundant codes, this type of noise can dominate the performance, even with suboptimal decoding (Pitkow et al. 2015). In this case, the remaining noise patterns *could* be averaged away by appropriate fine-scale weighting, but there is little benefit for the brain since the performance is limited by the redundancy. Inferring the actual fine-scale weighting pattern is therefore inconsequential compared to inferring the coarse-scale transformations that determine the information.

Here is one caveat, however: Although we extol the virtues of abstracting away from individual neuronal nonlinear mechanisms, nonetheless there may be certain functions that are difficult to implement as a combination of generic nonlinearities. For instance, both a quadratic nonlinearity and divisive normalization can be implemented as a sum of sigmoidally transformed inputs, but the latter requires a much larger number of neurons (Raju and Pitkow 2015). We speculate that cell types are hard-wired with specialized connectivity (Kim et al. 2014; Jiang et al. 2015) in order to accomplish useful operations, like divisive normalization, that are harder to learn by adjusting synapses between arbitrarily connected neurons.

*Messages, summary statistics, and neural responses*

Mathematically, message-passing algorithms operate on parameters of the graphical model. Yet in any practical implementation on a digital computer, the algorithms operate on elements — binary strings — that represent the underlying parameters. In our brain model, the corresponding elements are the spikes. The best way to understand machine computation may not be to examine the transformation of individual bits (Jonas and Körding 2017), but to look instead at the transformation of the variables those bits encode. Likewise, in the brain, we propose that it is more fundamental to describe the nonlinear transformation of encoded variables than to describe the detailed nonlinear response properties of individual neurons (Yang and Pitkow 2015; Raju and Pitkow 2016). The two nonlinearities may be related, but need not be, as we saw in Figure 2E.

Summary statistics provide a bridge between the representational and neural implementation levels. These statistics are functions of the neural activity that encode the posterior parameters. The updates to the posterior parameters define the algorithmic-level message-passing algorithm. The updates to the neural activity define a lower-level implementation, and will generally include activity patterns that are irrelevant to the computation. The updates to the summary statistics define the evolution of the subset of neural activity that actually represents the posterior parameters. It is these statistics that we describe in the article title as flowing in population codes (Figure 1).

*Behavioral modeling to estimate latent variables*

We expect that much of the brain's cortical machinery is dedicated to attributing dynamic latent causes to observations, and using them to choose appropriate actions. To understand this process, we need some experimental handle on the latent variables we expect the brain represents and transforms. Inferring latent variables requires perceptual models, and we measure an animal's percepts through its behavior, so we need behavioral models. We call attention to two types here.

The first behavioral model is a black box, such as an artificial recurrent neural network trained on a task. One can compare then the internal structure of the artificial network activity to the structure of real brain activity. If representational similarity (Kriegeskorte et al. 2008) between them suggests that the solutions are similar, one can then analyze the fully-observable artificial machinery to gain some insights into neural computation. This approach has been used fruitfully by (Mante et al. 2013; Yamins et al. 2014). However, such models are difficult to interpret without some external guess about the relevant latent variables and how they influence each other. This has been a persistent challenge even in artificial deep networks where we have arbitrary introspective power (Zeiler and Fergus 2014). In simple tasks, the relevant latent variables may be intuitively obvious. But in complex tasks we may not know how to interpret the computation beyond the similarity to the artificial network (Yamins et al. 2014), which makes it hard to understand and generalize. Ultimately, we need some principled way to characterize the latent variables.

This leads us to the second type of behavioral model: optimal control. Such a model uses probabilistic inference to identify the state of time-varying latent variables, their interactions, and the actions that maximize expected value. Clearly, animals are not globally optimal. Nonetheless, for some tasks, animals may understand the structure of the task, while mis-estimating its parameters and performing inference that is only approximate. Consequently, we can use the optimal structure to direct our search for computational features in neural networks, and fit the model parameters to best explain the actions we observe as based on evidence the animal observes. Fitting such a model is solving an inverse control problem (Ng and Russell 2000; Daunizeau et al 2010).

*Comparing behavioral and neural models*

A good behavioral model of this type provides us not only with parameters of the animal's internal model, but also with dynamic estimates of its beliefs and uncertainties about latent variables (Daw et al 2006; Kira et al. 2015) that evolve over time, perhaps through a message-passing inference algorithm. We can then examine how neuronal activity *encodes* these estimates at each time. The encodings can be fit by targeted dimensionality reduction methods ranging in complexity from linear regression to nonlinear neural networks. Once this encoding is fixed, we can estimate the agent's beliefs in two ways: one from the behavioral model, and one from the neural activity. Of course our estimates about the animal's estimates will be uncertain, and we will even have uncertainty about the animal's uncertainty. But if, up to an appropriate statistical confidence, we find agreement between these neurally derived quantities and those derived from the behavioral model, then this provides support that we have captured the encoding of statistics.

With an estimate of the neural encoding in hand, we should then evaluate whether the neural *recoding* matches the inference dynamics produced by the behavioral model. Recall that for a message-passing algorithm, the recoding process is a dynamical system that iteratively updates parameters of the posterior distribution over latent variables. In a neural implementation of this algorithm, recoding iteratively updates the neural activity through the recurrent connections, and the crucial aspect of these dynamics is how the summary statistics are updated: those statistics determine the posterior, so their updates determine the inference.

To measure the recoding observed in the brain, we recommend fitting a nonlinear dynamical system to the neural data. However, such systems can be extremely flexible, and would be pitifully underconstrained when fit to raw neural activity patterns. Thus computational models will benefit enormously from the dimensionality reduction of considering dynamics only on summary statistics about the task, not for all neural activities. Even so, we need hypotheses to constrain the dynamics. The message-passing framework provides one such hypothesis, by assuming the existence of canonical dynamics that operates over all latent variables. These dynamics are parameterized by two sets of parameters: one global set that defines a family of message-passing algorithms, and one task-specific set that specifies the graphical model over which the message-passing operates.

With a message-passing algorithm and underlying graphical model, both fit to explain the dynamics of neurally encoded summary statistics, we should then compare these sets of parameters between the behavioral and neurally derived models (on new data, of course). Agreement between these two recoding models is not guaranteed, even when the neural encoding was fit to the behavioral model. If the models disagree, then we can try to use the neurally-derived dynamics as a novel message-passing algorithm in the behavioral model. If we find agreement, then we have understood key aspects of neural recoding.

Once message-passing inference has localized probabilistic information to target variables, it can be used to select actions. But if an animal never guides any action by information in its neural populations, then it doesn't matter that neurons encode that information, or even if the network happens to transform it the right way. Thus it is critical to measure neural *decoding* — that is, how neural representations relate to behavior. Ideally, we would like to predict variations in behavior from variations in neural activity, especially from variations in those statistics that encode task-relevant parameters. A behavioral model predicts how latent variables determine actions. We can therefore test whether the trial-by-trial variations in the neural summary statistics predict future actions. If so, then we would have made major progress in understanding neural decoding.

The schematic in Figure 3 summarizes this approach of using a behavioral model with identifiable latent variables to interpret neural activity. In essence, this novel unified analysis framework allows one to perform hypothesis testing at the representational level, a step removed from the neural mechanisms, to measure the encoding, recoding, and decoding algorithms of the brain.

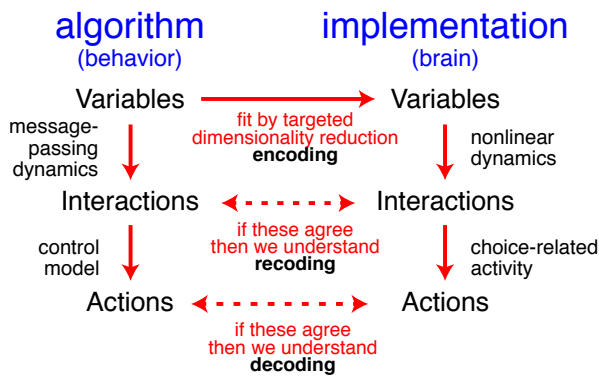

**Figure 3. Schematic for understanding distributed codes**
This figure sketches how behavioral models (left) that predict latent variables, interactions, and actions should be compared against neural representations of those same features (right). This comparison first finds a best-fit neural encoding of behaviorally relevant variables, and then checks whether the recoding and decoding based on those variables match between behavioral and neural models. The behavioral model defines latent variables needed for the task, and we measure the neural *encoding* by using targeted dimensionality reduction between neural activity and the latent variables (top). We quantify *recoding* of by the transformations of latent variables, and compare those transformations established by the behavioral model to the corresponding changes measured between the neural representations of those variables (middle row). Our favored model predicts that these interactions are described by a message-passing algorithm, which describes one particular class of transformations (Figure 1). Finally, we measure *decoding* by predicting actions, and compare predictions from the behavioral model and the corresponding neural representations (bottom row). A match between behavioral models and neural models, by measures like log-likelihood or variance explained, provides evidence that these models accurately reflect the encoding, recoding and decoding processes, and thus describe the core elements of neural computation.

*Task design to reveal flexible probabilistic computation*

Flexible, probabilistic nonlinear computation can best be revealed in concrete naturalistic tasks with interesting interactions. However, truly natural stimuli are too complex, and too filled with uncontrolled and indescribable nuisance variations, to make good computational theories (Rust and Movshon 2005). On the other hand, tasks should not be too simple, or else we won't be able to identify the computations. For a simple example, we saw in Figure 2D that the computations needed to infer angles were qualitatively different in the presence of nuisance variation, a more natural condition than many laboratory experiments. We want to understand the remarkable properties of the brain — especially those aspects that still go far beyond the piecewise-linear fitting of (impressively successful) deep networks (Krizhevsky et al. 2012). This means we need to challenge it to be flexible, to adjust processing dynamically. This requires us to find a happy medium: tasks that are neither too easy nor too hard.

Newer elaborations of 2AFC tasks require the animal to choose between multiple subtasks (Cohen and Newsome 2008; Mante et al 2013; Saez et al. 2015; Bondy and Cumming 2016). These are promising directions, as they provide greater opportunities for experiments to reveal flexible neural computation by introducing latent context variables. These contexts serve as nuisance variables that must be disentangled through nonlinear computation to infer the rewarded task-relevant variables. More nuisance variables will be better, but we should also be wary of natural tasks that are far too complex for us to understand (Rust and Movshon 2005).

To understand the richness of brain computations, the tasks we present to an animal should satisfy certain requirements. First, to understand nonlinear computations, one should include nuisance variables that the brain must untangle. Second, to reveal probabilistic inference, which hinges on appropriate treatment of uncertainty, one must manipulate uncertainty experimentally. Third, to expose an animal's internal model, the task should require the animal to predict unseen variables, for otherwise the animal can rely upon visible evidence which can compensate for any false beliefs an animal might harbor. Fourth, the task should be naturalistic, but neither too easy nor too hard. This has the best chances of keeping the brain engaged and the animal incentivized. These virtues would address the problems arising from overly simple tasks and would help us refine our understanding of the neural basis of behavior.

## CONCLUSION

In this paper, we proposed that the brain naturally performs approximate probabilistic inference, and critiqued overly simple tasks as being ill suited to expose the inferential computations that make the brain special. We introduced a hypothesis about computation in cortical circuits. Our hypothesis has three parts. First, overlapping patterns of population activity encode statistics that summarize both estimates and uncertainties about latent variables. Second, the brain specifies how those variables are related through a sparse probabilistic graphical model of the world. Third, recurrent circuitry implements a nonlinear message-passing algorithm that selects and localizes the brain's statistical summaries of latent variables, so that all task-relevant information is actionable.

Finally, we emphasized the advantage in studying the computation at the level of neural population activity, rather than at the level of single neurons or membrane potentials: If the brain does use redundant population codes, then many fine details of neural processing don't matter for computation. Instead it can be beneficial to characterize computation at a more abstract representational level, operating on variables encoded by populations, rather than on the substrate.

Naturally, understanding the brain is difficult. Even with the torrents of data that may be collected in the near future, it will be challenging to build models that reflect meaningful neural computation. A model of approximate probabilistic inference operating at the representational level will be an invaluable abstraction.

## ACKNOWLEDGEMENTS

The authors thank Jeff Beck, Greg DeAngelis, Ralf Haefner, Kaushik Lakshminarasimhan, Ankit Patel, Alexandre Pouget, Paul Schrater, Andreas Tolias, Rajkumar Vasudeva Raju, and Qianli Yang for valuable discussions. XP was supported in part by a grant from the McNair Foundation, NSF CAREER Award IOS-1552868, a generous gift from Britton Sanderford, and by the Intelligence Advanced Research Projects Activity (IARPA) via Department of Interior/Interior Business Center (DoI/IBC) contract number D16PC00003.[1] XP and DA are supported in part by the Simons Collaboration on the Global Brain award 324143, and BRAIN Initiative grants NIH 5U01NS094368 and NSF 1450923 BRAIN 43092-N1.

---